\documentclass[12pt]{article}
\usepackage{graphicx}
\usepackage{amsmath}
\usepackage{authblk}
\usepackage{hyperref}
\usepackage[margin=1in]{geometry}
\usepackage{booktabs}
\usepackage{tabularx}
\usepackage{multirow}
\usepackage{siunitx}
\usepackage{abstract}

\title{ProtT-Affinity: Sequence-Based Protein--Protein Binding Affinity Prediction Using ProtT5 Embeddings}

\author[1]{Hongfu Lou}

\affil[1]{Department of Physics, Huazhong University of Science and Technology}

\date{}

\begin{document}
\maketitle

Author Contributions: H.L. conceived the study, designed the methodology, performed all experiments, analyzed the data, and wrote the manuscript.

Competing Interests:The author declares no competing interests.

\begin{abstract}
\noindent
\parindent=0pt
Predicting the binding affinity of protein–protein complexes directly from sequence remains a challenging problem, particularly in the absence of reliable structural information. Here I present \textbf{ProtT-Affinity}, a sequence-only model that combines ProtT5 embeddings with a lightweight Transformer architecture. The model is trained and evaluated on homology-filtered subsets of the PDBBind database following a curation protocol consistent with prior structure-based work. Across two independent test sets, ProtT-Affinity reaches Pearson correlation coefficients of 0.628 and 0.459, respectively. Although its performance does not match the strongest structure-based methods, it is competitive with several widely used approaches and provides a practical alternative when structural data are missing or uncertain. The results suggest that large protein language models capture features relevant to binding energetics, and that these features can be exploited to approximate affinity trends at scale. All code and data used in this study are available at \url{https://github.com/fufuhonghong/ProtT-Affinity/}
.
\end{abstract}

\textbf{Keywords:} Protein--protein interaction, Binding affinity prediction, Sequence-based model

\section{Introduction}

Protein–protein interactions (PPIs) underpin a wide range of cellular processes, including signal transduction, immune recognition, and transcriptional regulation\cite{sun2017sequenceppi}. Characterizing the binding affinity of protein complexes is therefore central to understanding biological mechanisms and plays an important role in areas such as drug discovery and protein engineering\cite{zhou2024proaffinity}. Experimental measurements, however, remain time-consuming and often difficult to scale, which has motivated the development of computational alternatives.

A large body of work has approached affinity prediction from a structural perspective. Methods such as DFIRE\cite{liu2004physical}, PRODIGY\cite{xue2016prodigy}, and the more recent ProAffinity-GNN\cite{zhou2024proaffinity} make explicit use of high-resolution complex structures, drawing on energy-based or graph-based representations. Other efforts, including AFTGAN\cite{kang2023aftgan} and PLANET\cite{zhang2024planet}, have incorporated attention mechanisms and advanced neural architectures to improve predictive performance. While these approaches have demonstrated strong accuracy, their dependence on experimentally solved or confidently predicted structures limits their applicability to large-scale sequence datasets.

Sequence-based approaches offer a more flexible alternative. Early methods relied mainly on handcrafted features or traditional machine learning models\cite{guo2010predppi,gui2020dnnppi,mahapatra2022dnnxgb}. Subsequent deep learning models such as DNN-PPI\cite{zhang2018ensemble} and PIPR\cite{kozakov2006piper} showed improved generalization by learning patterns directly from sequence. More recent studies have leveraged protein language models (PLMs)—for instance BAPULM\cite{meda2024bapulm} and PPIretrieval\cite{hua2024ppiretrieval}—which capture evolutionary and functional relationships without requiring multiple sequence alignments or structural templates.

Building on these developments, this work introduces \textbf{ProtT-Affinity}, a sequence-only framework for predicting the binding affinity of protein–protein complexes. The approach employs ProtT5\cite{elnggar2022prottrans} embeddings in combination with a compact Transformer encoder to model interactions between partner sequences. The model is evaluated under the same stringent data curation protocol used by structure-based benchmarks\cite{zhou2024proaffinity}, enabling a more direct comparison between sequence-derived and structure-derived representations. In doing so, this study also explores the broader question of how effectively large PLMs encode affinity-relevant properties and whether such embeddings can approximate structural information for quantitative tasks.

\section{Methods}

\subsection{Dataset Preparation and Curation}

The dataset used in this study follows the curation procedure described by Zhou et al.\cite{zhou2024proaffinity} and is based on the protein–protein complex subset of the PDBBind 2020 release\cite{wang2004pdbbind}. Starting from 2,852 complexes with experimentally measured binding affinities, complexes involving nucleic acids were removed, and only entries with reported dissociation constants ($K_d$) were retained, resulting in 2,071 usable samples.

To reduce redundancy, the dataset was filtered using a sequence identity cutoff of 25\%, following the homology reduction protocol in\cite{zhou2024proaffinity}. This produced a final training set of 1,741 complexes. Two separate test sets were adopted from the same reference: (i) 79 complexes from a commonly used structure-based benchmark\cite{kastritis2011structure}, and (ii) 82 two-chain complexes drawn from PDBBind. Consistent with previous affinity prediction studies\cite{zhou2024proaffinity,zhang2024planet}, all $K_d$ values were converted to $pK_a = -\log_{10}(K_d)$ prior to training and evaluation.

\subsection{Sequence Representation and Feature Extraction}

Protein sequences were represented using embeddings generated by the ProtT5-XL-U50 model\cite{elnggar2022prottrans}, a transformer-based protein language model trained on hundreds of millions of sequences. ProtT5 adopts the T5 architecture originally developed for natural language processing, but is pre-trained on large-scale protein databases using a masked language modeling objective. Through this pre-training strategy, the model learns contextual relationships among amino acids in a way that reflects both evolutionary conservation and higher-level functional constraints.

Several recent studies have shown that embeddings derived from ProtT5 implicitly capture a broad range of structural features, including secondary structure tendencies, disorder, and aspects of long-range residue interactions. These properties make ProtT5 a suitable backbone for tasks that traditionally rely on structural information but for which explicit 3D models may be unavailable or unreliable.

In this work, each protein sequence was passed through ProtT5 to obtain a sequence of 1,024-dimensional residue embeddings. To construct a fixed-size representation for downstream modeling, we averaged these residue-level vectors to produce a single embedding for each protein. Although this pooling operation discards position-specific details, we found it sufficient for capturing overall biophysical trends relevant to binding energetics. The embeddings of the two interacting proteins were concatenated into a 2,048-dimensional vector, providing the model with a combined representation of the pair. This setup aims to retain complementary information from both partners while keeping the overall architecture lightweight and computationally manageable.

\subsection{Model Architecture}

\begin{figure}[ht]
\centering
\includegraphics[width=1.0\textwidth]{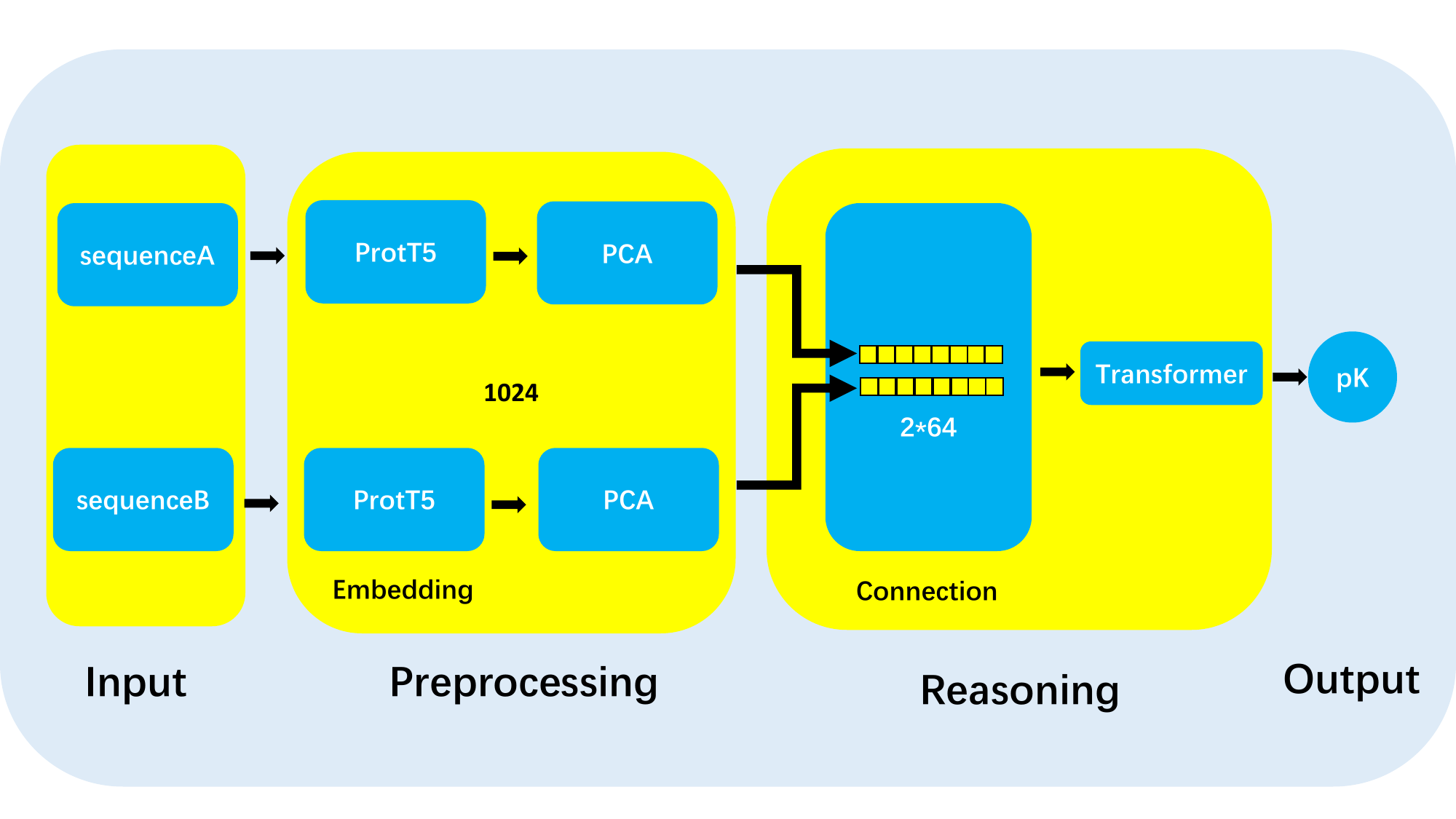}
\caption{Architecture of the ProtT-Affinity model. ProtT5 embeddings of both interacting proteins are processed through a Transformer encoder and regression head to predict the binding affinity.}
\label{fig:architecture}
\end{figure}

The overall model design is illustrated in Figure~\ref{fig:architecture}. ProtT5 embeddings for each protein serve as the initial input. In some experiments, we applied Principal Component Analysis (PCA) to reduce dimensionality, mainly to assess whether a more compact representation would affect performance. Inter-protein dependencies were modeled using a bidirectional cross-attention module: one attention block computes the influence of protein A on protein B, and a second block computes the reverse. Each block employs four attention heads, which allows the model to capture a range of interaction patterns.

Outputs from the cross-attention layers were averaged along the sequence dimension and concatenated, yielding a joint representation for the paired proteins. This vector was passed through a small feed-forward network with ReLU activations and dropout for regularization, producing the final $pK_a$ prediction. Source code and processed datasets are available at: \url{https://github.com/fufuhonghong/ProtT-Affinity/}
.

\subsection{Training Protocol and Evaluation Metrics}

The model was trained using the Huber loss ($\delta = 2.0$), which we found to be more stable than mean squared error when dealing with occasional large deviations. Optimization was performed using AdamW with a learning rate of $1\times10^{-4}$ and weight decay of the same magnitude. Training ran for up to 200 epochs with early stopping (patience of 15 epochs) based on validation loss. Batch size was set to 16, and gradients were clipped at a norm of 1.0 to avoid instability.

A ReduceLROnPlateau scheduler (factor 0.5, patience 8) was used to adjust the learning rate when validation performance plateaued. All experiments were repeated with three random seeds, and reported values correspond to the mean and standard deviation across these runs. Performance was evaluated using Pearson’s correlation coefficient (R), mean absolute error (MAE), mean squared error (MSE), and the coefficient of determination ($R^2$).

\begin{table}[ht]
\centering
\caption{Training hyperparameters and settings.}
\label{tab:training_params}
\begin{tabular}{ll}
\hline
Parameter & Value \\
\hline
Loss function & Huber loss ($\delta=2.0$) \\
Optimizer & AdamW \\
Learning rate & $1\times10^{-4}$ \\
Weight decay & $1\times10^{-4}$ \\
Batch size & 16 \\
Maximum epochs & 200 \\
Early stopping patience & 15 \\
Gradient clipping norm & 1.0 \\
Learning rate scheduler & ReduceLROnPlateau \\
Scheduler factor & 0.5 \\
Scheduler patience & 8 \\
Sample weighting & Based on deviation from mean pK \\
\hline
\end{tabular}
\end{table}

\section{Results}

\subsection{Performance Evaluation}

The predictive performance of ProtT-Affinity on the two benchmark test sets is summarized in Table~\ref{tab:results}. On the first test set, which consists primarily of complexes drawn from a widely used structure-based affinity benchmark, the model achieved a Pearson correlation coefficient of 0.628 and a mean absolute error (MAE) of 1.645 kcal/mol. Performance on the second test set was somewhat lower (R = 0.459; MAE = 1.794 kcal/mol), which is consistent with prior reports indicating that this subset includes a broader diversity of complex types and generally poses a more challenging prediction task.

When the two test sets are combined, ProtT-Affinity obtains an overall correlation of 0.579 and an MAE of 1.722 kcal/mol. While these values do not reach the accuracy of the best structure-dependent methods, they demonstrate that sequence-only features extracted from protein language models capture a substantial portion of the variation in experimental binding affinities. Importantly, the results were stable across the three independent training runs, with standard deviations remaining small relative to the absolute performance metrics.

\begin{table}[ht]
\centering
\caption{Performance of ProtT-Affinity on PDBBind benchmark sets (mean ± std over three runs).}
\begin{tabular}{lccc}
\toprule
Dataset & Size & Pearson's R & MAE (kcal/mol) \\
\midrule
Test set 1 & 79 & 0.628 ± 0.015 & 1.645 ± 0.032 \\
Test set 2 & 82 & 0.459 ± 0.021 & 1.794 ± 0.028 \\
Combined set & 161 & 0.579 ± 0.018 & 1.722 ± 0.030 \\
\bottomrule
\end{tabular}
\label{tab:results}
\end{table}

\subsection{Comparison with Existing Methods}

To contextualize the performance of ProtT-Affinity, we compared the model with a range of sequence-based and structure-based predictors under the same benchmarking protocol (Table~\ref{tab:performance_benchmarks}). As expected, state-of-the-art structure-derived models such as ProAffinity-GNN outperform sequence-only approaches, particularly on the more heterogeneous test sets. Nevertheless, ProtT-Affinity achieves comparable accuracy to the traditional $\text{PPI-Affinity SVM}$ model and surpasses several well-established baselines, including $\text{DFIRE}$ and $\text{CP\_PIE}$.

Interestingly, the gap between sequence-based and structure-based methods is not uniform across datasets. On Test Set 1, ProtT-Affinity is noticeably closer to the performance of mid-tier structure-based methods, whereas on Test Set 2 the difference widens. This may reflect the extent to which the underlying complexes conform to the types of interactions that PLM embeddings can represent effectively. Several recent studies have noted that PLM-derived features tend to correlate well with general trends in binding energetics but may struggle when fine-grained structural details dominate the interaction landscape.

\begin{figure}[ht]
\centering
\includegraphics[width=0.95\textwidth]{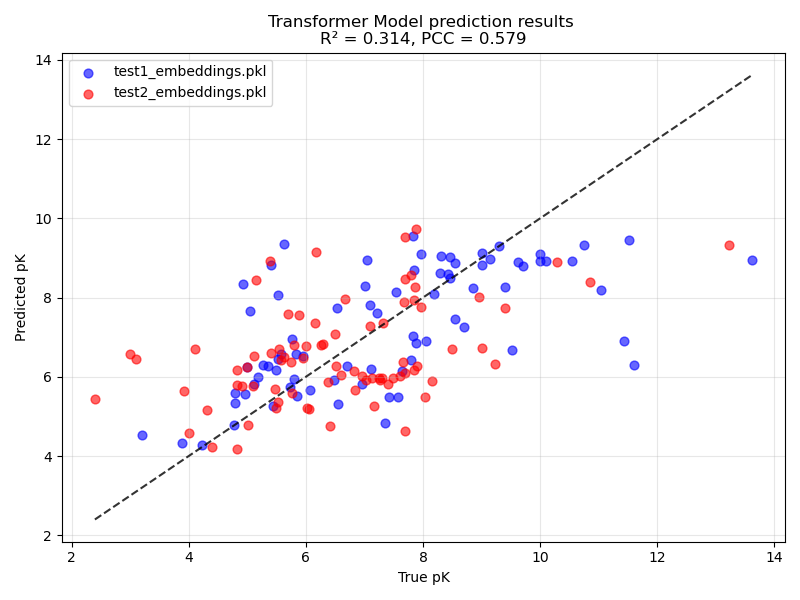}
\caption{Correlation between predicted and experimental $pK_a$ values across test sets. (A) Test set 1 (R = 0.628), (B) Test set 2 (R = 0.459). The diagonal indicates perfect agreement. }
\label{fig:correlation}
\end{figure}

Figure~\ref{fig:correlation} illustrates the relationship between predicted and experimental $pK_a$ values for both test sets. The scatterplots show that the model captures the overall trends in affinity but exhibits larger variability for complexes with intermediate binding strengths. This is consistent with the behavior of many sequence-based models and suggests that hybrid approaches incorporating structural priors or predicted complex geometries may further improve performance.
\begin{table}[ht]
\small
\caption{Comprehensive comparison with existing methods on standard benchmarks.}
\begin{tabular}{lccccccc}
\toprule
\multirow{2}{*}{Method} & \multirow{2}{*}{Type} & \multicolumn{2}{c}{Test set 1 (79)} & \multicolumn{2}{c}{Test set 2 (82)} & \multicolumn{2}{c}{Combined set (161)} \\
\cmidrule(r){3-4} \cmidrule(r){5-6} \cmidrule(r){7-8}
 & & R & MAE & R & MAE & R & MAE \\
\midrule
PRODIGY\cite{zhou2024proaffinity,xue2016prodigy} & Structure & 0.735 & 1.43 & 0.334 & 2.52 & 0.456 & 1.98 \\
DFIRE\cite{zhou2024proaffinity,liu2004physical} & Structure & 0.602 & 4.64 & 0.145 & 26.02 & -0.005 & 15.53 \\
CP\_PIE\cite{zhou2024proaffinity,ravikant2010pie} & Structure & 0.517 & 8.80 & 0.111 & 7.26 & 0.167 & 8.02 \\
ISLAND\cite{zhou2024proaffinity,abbasi2020island} & Sequence & 0.378 & 2.10 & 0.217 & 2.15 & 0.314 & 2.13 \\
PPI-Affinity (SVM)\cite{zhou2024proaffinity,romero2022ppi} & Structure & 0.616 & 1.82 & 0.436 & 1.78 & 0.545 & 1.80 \\
ProAffinity-GNN\cite{zhou2024proaffinity} & Structure & 0.697 & 1.52 & 0.620 & 1.49 & 0.669 & 1.50 \\
\textbf{ProtT-Affinity } & \textbf{Sequence} & \textbf{0.628} & \textbf{1.65} & \textbf{0.459} & \textbf{1.79} & \textbf{0.579} & \textbf{1.72} \\
\bottomrule
\end{tabular}
\label{tab:performance_benchmarks}
\end{table}

These findings are consistent with recent sequence-only studies such as BAPULM\cite{meda2024bapulm} and PPIretrieval\cite{hua2024ppiretrieval}, which confirm that PLMs can effectively capture binding-related representations without explicit structure input.

\section{Discussion}

ProtT-Affinity establishes that large-scale protein language models can effectively infer binding energetics solely from sequence. By leveraging ProtT5 embeddings and attention mechanisms, it achieves performance competitive with sophisticated structure-dependent predictors. These results echo recent findings that PLMs implicitly encode secondary and tertiary structure features\cite{soleymani2022ppireview}.

In pharmaceutical research, the need to screen massive protein libraries for drug development is often constrained by the high computational costs of structure-based methods. ProtT-Affinity, by leveraging sequence-only inference, avoids the resource-heavy demands of structural modeling (such as molecular docking or 3D structure prediction), thus drastically reducing computational overhead. This efficiency enables rapid prioritization of promising protein targets, accelerating the transition from target discovery to experimental validation and ultimately shortening the timeline of drug development pipelines.

Despite its strengths, ProtT-Affinity remains limited in residue-level interpretability and cannot yet rival structure-based approaches such as ProAffinity-GNN in absolute accuracy. Future research could integrate predicted 3D structures from AlphaFold, or hybridize graph-based frameworks such as PPI-Graphomer\cite{xie2025ppigraphomer} and AFTGAN\cite{kang2023aftgan} for enhanced contextual reasoning. Expanding training datasets following strategies from PLANET\cite{zhang2024planet} or Lv et al.\cite{lv2021gnnppi} may further improve generalization across unseen protein families.

\section{Conclusion}

This work introduces \textbf{ProtT-Affinity}, a scalable, sequence-only framework for protein–protein binding affinity prediction. Through rigorous benchmarking, this work demonstrates that sequence embeddings from large PLMs can approximate the predictive power of structure-based models, offering an accessible alternative when 3D information is unavailable. This work advances the frontier of sequence-driven biophysical modeling and lays a foundation for future integration with structure prediction and interaction analysis tools.

The success of ProtT-Affinity exemplifies the growing synergy between natural language processing and biophysics, marking a step toward universal, structure-free modeling of molecular interactions.

\bibliographystyle{unsrt}
\bibliography{references}

\end{document}